\documentclass[
    aps,
    prl,
    reprint,
    superscriptaddress,
    amsmath,
    amssymb,
    nofootinbib,
]{revtex4-2}

\usepackage{newtxtext}
\usepackage{newtxmath}

\usepackage{mathtools}   
\usepackage{bm}          
\usepackage{dsfont}    

\usepackage{graphicx}
\usepackage{booktabs}
\usepackage{multirow}
\usepackage{makecell}

\usepackage{microtype}

\usepackage[
    colorlinks=true,
    linkcolor=blue,
    citecolor=blue,
    urlcolor=blue
]{hyperref}

\usepackage{physics}
\usepackage{rotating}
\usepackage{makecell}

\DeclareMathOperator{\sso}{SSO}

\DeclareMathOperator{\phasegate}{Phase_{000}}

\newcommand{\ident}{\mathds{1}}

\newcommand{\nitrogen}{${}^{14}\textrm{N}$}
\newcommand{\nitrogenmath}{{}^{14}\textrm{N}}
\newcommand{\carbon}{${}^{13}\textrm{C}$}
\newcommand{\carbonmath}{{}^{13}\textrm{C}}

\newcommand{\intensityvector}{\vec{I}}
\newcommand{\propabilityvector}{\vec{p}}

\newcommand{\kete}[1]{\ket{#1}_{e}}

\newcommand{\oracleop}{U}
\newcommand{\reflectionop}{R}
\newcommand{\groverop}{G}

\begin{document}

\title{Grover Search with Semiconductor Spin Qubits at Ambient Conditions}


\author{Sebastian Gemsheim}
\affiliation{
    SAXON Q GmbH,
    Emilienstra\ss e 15,
    04107 Leipzig, Germany
}

\author{Fabian Klüpfel}
\affiliation{
    SAXON Q GmbH,
    Emilienstra\ss e 15,
    04107 Leipzig, Germany
}

\author{Max Knei\ss}
\affiliation{
    SAXON Q GmbH,
    Emilienstra\ss e 15,
    04107 Leipzig, Germany
}

\author{Nicole Raatz}
\affiliation{
    SAXON Q GmbH,
    Emilienstra\ss e 15,
    04107 Leipzig, Germany
}

\author{Tobias Herzig}
\affiliation{
    SAXON Q GmbH,
    Emilienstra\ss e 15,
    04107 Leipzig, Germany
}

\author{Matthias Mendt}
\affiliation{
    SAXON Q GmbH,
    Emilienstra\ss e 15,
    04107 Leipzig, Germany
}

\author{Evgeny Kreissig}
\affiliation{
    SAXON Q GmbH,
    Emilienstra\ss e 15,
    04107 Leipzig, Germany
}

\author{Ulrike R\"uckert}
\affiliation{
    Fraunhofer IWU, 
    N\"othnitzer Stra\ss e 44,
    01187 Dresden, Germany
}

\author{Albrecht H\"anel}
\affiliation{
    Fraunhofer IWU, 
    N\"othnitzer Stra\ss e 44,
    01187 Dresden, Germany
}

\author{Jan Meijer}
\affiliation{
    SAXON Q GmbH,
    Emilienstra\ss e 15,
    04107 Leipzig, Germany
}
\affiliation{
  Felix-Bloch-Institute for Solid State Physics, 
  University of Leipzig, 
  Linn\'estra\ss e 5,
  04103 Leipzig, Germany
}

\author{Marius Grundmann}
\affiliation{
    SAXON Q GmbH,
    Emilienstra\ss e 15,
    04107 Leipzig, Germany
}
\affiliation{
  Felix-Bloch-Institute for Solid State Physics, 
  University of Leipzig, 
  Linn\'estra\ss e 5,
  04103 Leipzig, Germany
}


\begin{abstract}
Grover’s algorithm is executed on a commercial quantum computer based on nitrogen-vacancy centers in diamond operating under ambient conditions, achieving fidelities up to 99.98\,\%. Within a $N=8$ search space of three solid-state nuclear spin qubits, the measured success probabilities of finding a single or two marked states are $(77.3 \pm 3.4)\,\%$ and $(87.0 \pm 4.2)\,\%$, respectively. These values surpass published results for superconducting qubits or any quantum computer operating at room temperature. In addition, the paper also details the calibrated fidelities of the implemented universal gate set.
\end{abstract}

\maketitle


\emph{Introduction.---}
Information processing in the quantum realm provides opportunities for algorithmic advantages over classical computing. Grover search~\cite{Grover1996} in particular, either as a stand-alone search algorithm or a subroutine for other quantum algorithms, has been one of the early discoveries in quantum information theory exhibiting better-than-classical scaling. Numerous experimental demonstrations thereof have been reported for different modalities such as neutral atoms~\cite{Ahn2000}, liquid-state nuclear magnetic resonance~\cite{Chuang1998, Vandersypen2000}, superconducting devices~\cite{DiCarlo2009, Roy2020, Zhang2021, Zhang2022, Chu2022, Roy2023, Pokharel2024, Liu2025, Leng2025}, trapped ions~\cite{Figgatt2017, Zhang2022, Main2025, Butt2026, Shi2026}, photonic platforms~\cite{Ciampini2016, Feng2026, Huster2026} and silicon spin qubits~\cite{Watson2018, Noiri2022, Thorvaldson2025}. 

Each of the mentioned modalities requires different environmental conditions. Superconducting devices and silicon spin qubits solely work at ultralow temperatures in a vacuum environment. Further, trapped ion (and neutral atom) hardware can only operate in ultra-high vacuum and require laser cooling for a reduction of atomic motion. Lastly, even though photonic processors can operate at room temperature, their high-efficiency single photon detection necessitates cryogenic temperatures.

Nitrogen-vacancy (NV) centers in diamond~\cite{Pezzagna2021} offer an alternative route for the implementation of quantum algorithms. The diamond lattice provides the two-fold advantage of providing a robust environment and supplying additional nuclear carbon isotope spins as qubits. Both properties enabled pioneering demonstrations of quantum information processing~\cite{Waldherr2014, Taminiau2014, Reiserer2016, Kalb2017, Bradley2019, Pezzagna2021, Abobeih2022, Iuliano2026}, and particularly Grover searches~\cite{vanderSar2012, Wu2019, Zhang2020} with two qubits at ambient conditions. These algorithmic benchmarks are essential for evaluating the correct interplay of all components and the operational performance of a quantum device as a whole. 

In this work, we perform a three-qubit quantum search on a commercially available room temperature quantum computer based on diamond, installed on the premises of Fraunhofer IWU. As a first demonstration of this size with NV center technology, our results are on par with Grover performances from other modalities. Different from previous work, our quantum computation is solely based on nuclear spin qubits and does not need to employ the electronic NV center spin as a computational unit. This approach leverages the temperature stability and excellent coherence properties ($T_2^* \sim \text{1--10\,ms}$)~\cite{Maurer2012} of hyperfine-coupled surrounding nuclear spins, which are roughly three orders of magnitude larger than the typical electronic spin coherence times.

In the following, individual benchmark results of all native gates precede the presentation of our algorithmic results. The rest of the Letter is designated for the hardware calibration procedure enabling the aforementioned results and for future improvements.
\\

\emph{Benchmarking of native gate set.---}
On our device, the electronic spin ($S=1$) of the NV center provides an optical interface for readout and facilitates the nuclear entangling operation. In addition, it functions as intermediary for the full polarization of the relevant nuclear spins.
The set of computational qubits is composed of the inherent nitrogen \nitrogen{} spin ($I=1$) of the NV center and two strongly coupled, surrounding \carbon{} carbon spins ($I=1/2$). Specifically, only a two-level subspace of the three-level \nitrogen{} system is used for computation. Each nuclear spin can be addressed through radiofrequency pulses, which, together with virtual $z$-rotations~\cite{McKay2017}, implement arbitrary single-qubit unitary gates. 
A multi-qubit phase gate $\phasegate = \ident - 2\dyad{000}$, locally equivalent to a Toffoli gate, not only completes the native gate set, but renders it universal. Its realization is based on a geometric phase induced through selective microwave pulses on the electronic spin of the NV center. As such, it is rather a three-controlled four-qubit gate, but we omit this notation for the purpose of focusing on the computational state space.
Our working principle is similar to the one employed in Ref.~\cite{Waldherr2014}.

\begin{figure}
  \centering
  \includegraphics[width=0.49\columnwidth]{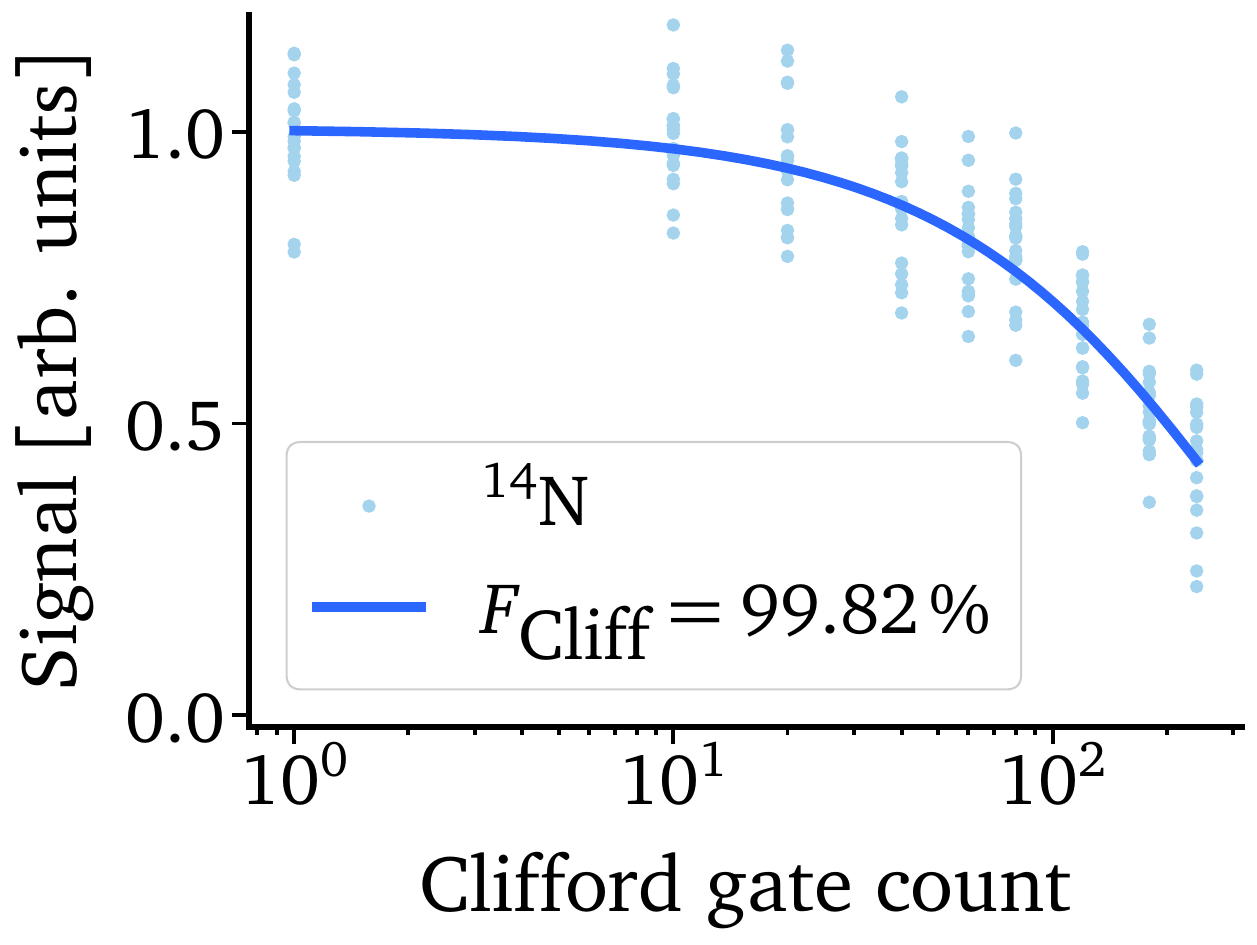}
  \includegraphics[width=0.49\columnwidth]{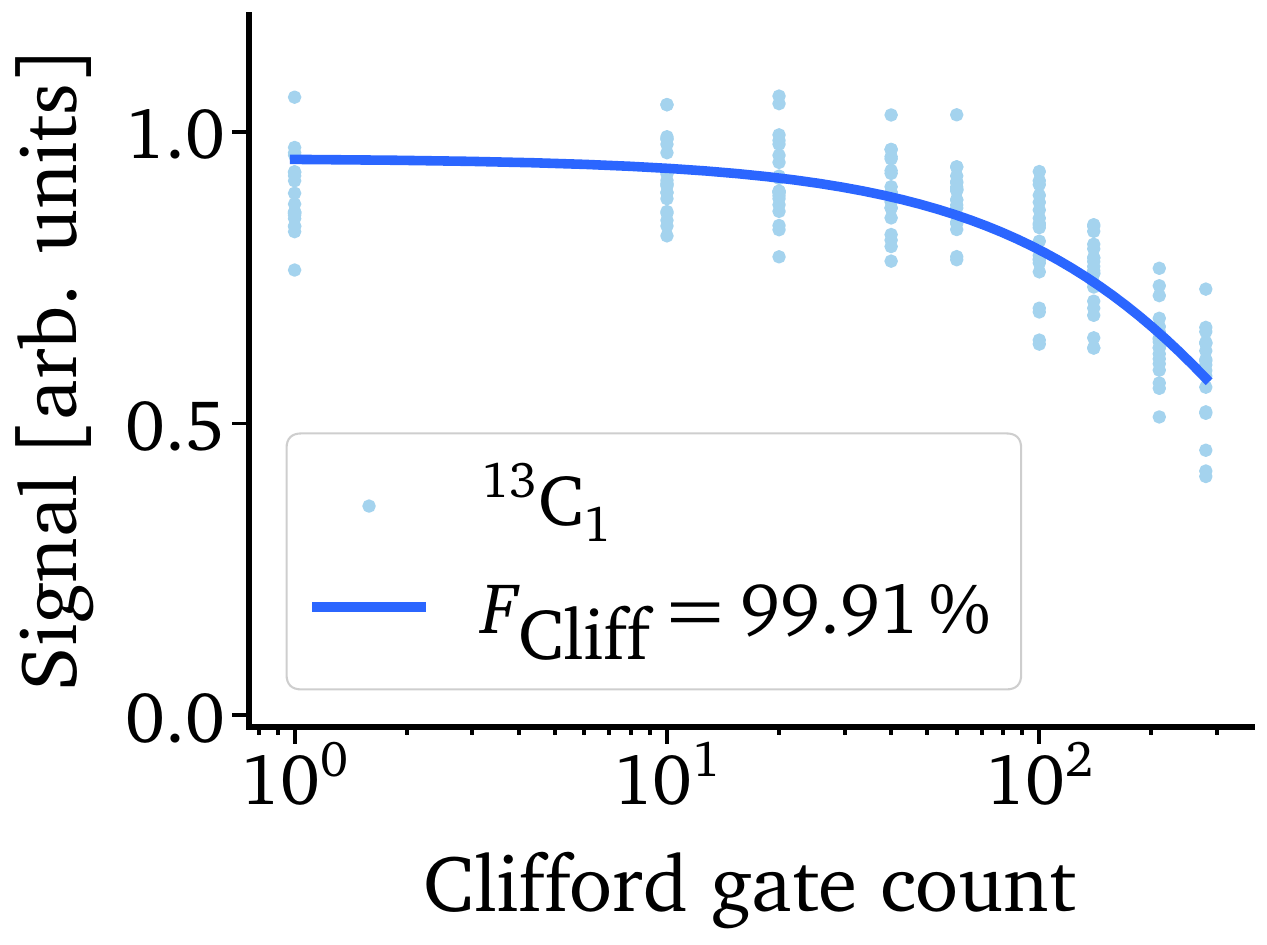}
  \\
  \hfill
  \includegraphics[width=0.49\columnwidth]{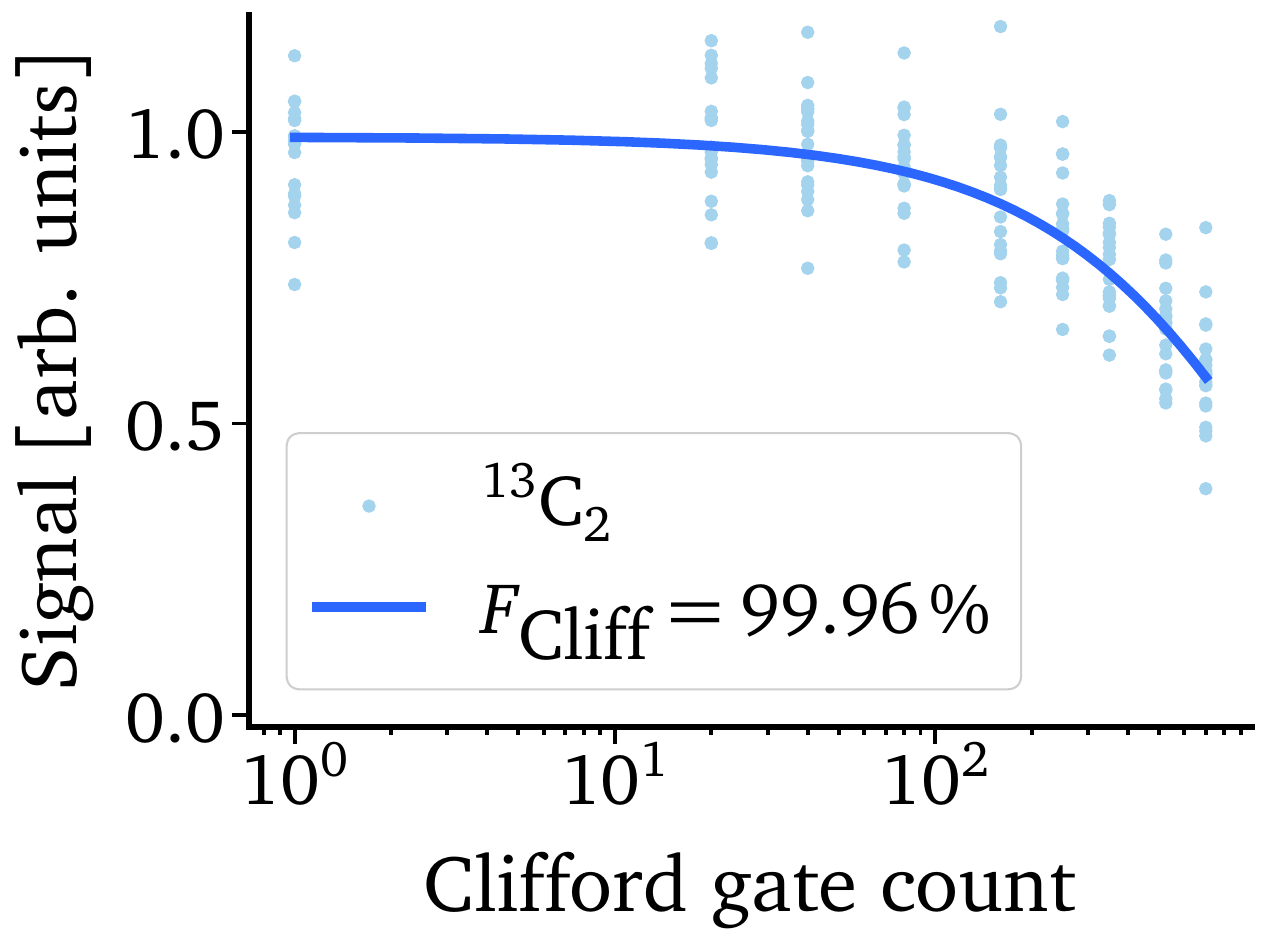}
  \includegraphics[width=0.49\columnwidth]{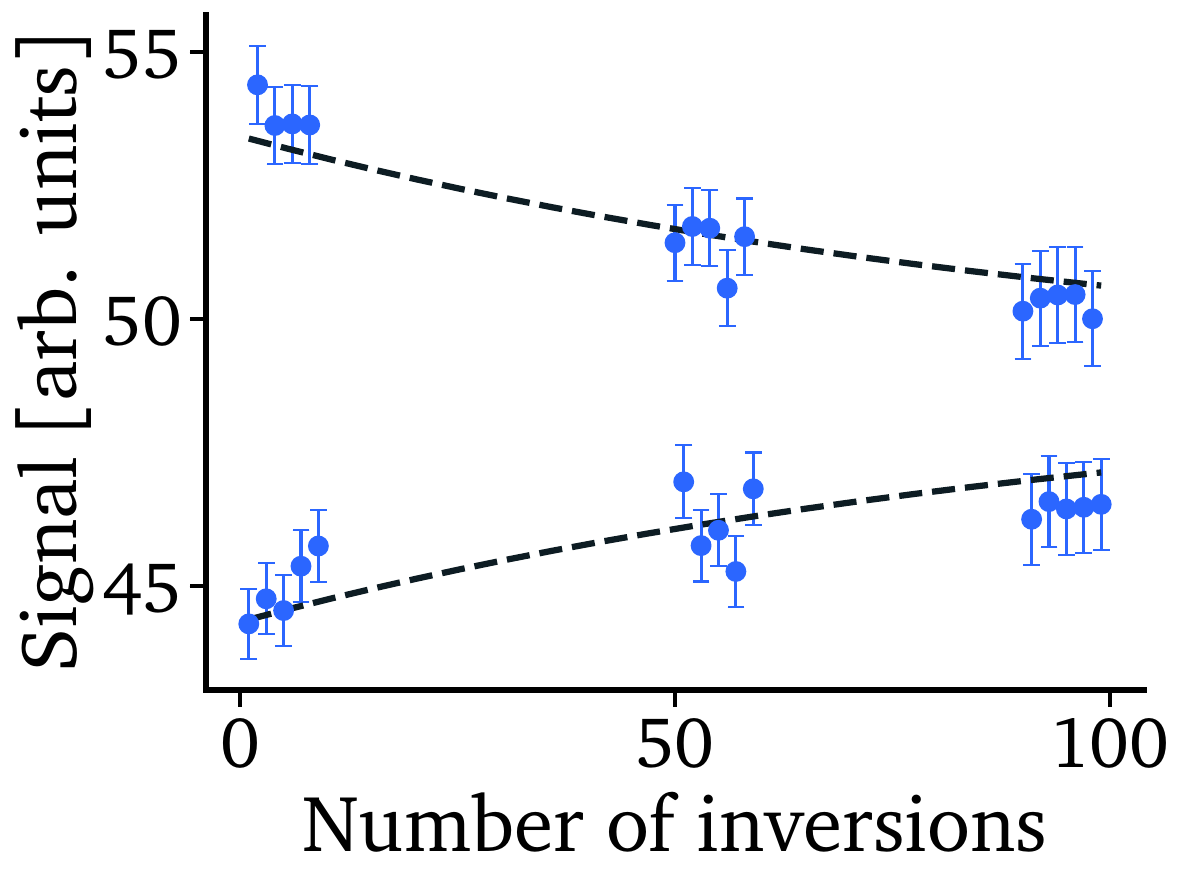}
  \caption{
      Single qubit randomized benchmarking results and electron spin inversion performance. For each number $n$ of Clifford gates in the RB sequence, we randomly sample 20 different RB circuits and fit the overall result by $f(n) = A p^n + B$. While $A$ and $B$ capture SPAM errors, the effective depolarizing parameter $p$ is directly related to the average fidelity through $F = (1+p)/2$ for single qubits~\cite{Hashim2025}.
      The corresponding average Clifford gate fidelities are $F_{\text{Cliff}}[\nitrogenmath] = (99.82 \pm 0.07)\,\%$, $F_{\text{Cliff}}[\carbonmath_{1}] = (99.91 \pm 0.08)\,\%$ and $F_{\text{Cliff}}[\carbonmath_{2}] = (99.96 \pm 0.03)\,\%$. 
      Additionally, the fidelities for a native $S_x$ (equivalent to a rotation $R_x(\pi/2)$ around the $x$-axis on the Bloch sphere) derive from the Clifford gate error rates and yield $F_{S_x}[\nitrogenmath] = (99.91 \pm 0.04)\,\%$, $F_{S_x}[\carbonmath_{1}] = (99.95 \pm 0.04)\,\%$ and $F_{S_x}[\carbonmath_{2}] = (99.98 \pm 0.02)\,\%$.
      The lower right plot shows repetitive applications of the robust inversion pulses for the electronic spin. From an exponential fit (dashed lines), we determine a pseudo-fidelity of 99.1\,\%. 
  }
  \label{fig:1q_rb}
\end{figure}

While a full process tomography would reveal details about the specific errors occuring during gate executions, we opt for randomized benchmarking (RB)~\cite{Hashim2025} as means for obtaining \emph{average} gate fidelities independent of state preparation and measurement (SPAM) errors. For each of the nuclear spin qubits, the benchmark outcomes are shown in Figure~\ref{fig:1q_rb}. Based on these results, we find an average Clifford gate fidelity of $99.90\,\%$ and an average $S_x$ gate fidelity of $99.95\,\%$.

For larger gate dimensions, the simplicity of standard RB comes at the cost of significant resource overheads and necessitates alternative benchmarking methods.
To this end, we use a modified version~\cite{Dubovitskii2022} of interleaved randomized benchmarking (IRB)~\cite{Hashim2025} with single-qubit Clifford twirls to estimate two-qubit-subspace fidelities $F_{q0,q1} = (96.0 \pm 0.3)\,\%$, $F_{q0,q2} = (96.5 \pm 0.3)\,\%$ and $F_{q1,q2} = (94.8 \pm 0.3)\,\%$ for the $\phasegate$ gate, or $95.7\,\%$ on average.
Taken together, this native gate set possesses a sufficiently high quality for the implementation of a search algorithm.
\\

\emph{Grover search---}
Searching for a specific item in an unsorted database of $N$ total elements requires roughly $\mathcal{O}(N)$ steps through classical brute force until solution. In contrast, the now well-known quantum Grover search~\cite{Nielsen2000} can achieve this feat in $\mathcal{O}(\sqrt{N})$ queries, a quadratic speedup. 
At the core, both in the classical and quantum case, a solution candidate $x$ (integer or bitstring) gets classified via the query of an oracle function
\begin{align}
  f(x) = \begin{cases}
    1 & x\text{ is solution} \\
    0 & \text{otherwise} 
  \end{cases} \,,
\end{align}
which indicates if $x$ is a valid solution or not. A corresponding quantum oracle $\oracleop$ acts as $\oracleop\ket{x} = (-1)^{f(x)} \ket{x}$ on any basis state $\ket{x}$ and assigns a minus sign to valid solutions. 
Starting from an equally-weighted superposition state $\ket{\psi} = H^{\otimes 3} \ket{000}$ for our three qubits with $N=2^3$, in each Grover step the oracle-tagged solution amplitudes are amplified through the reflection operator $\reflectionop = \ident - 2 \dyad{\psi} = H^{\otimes 3} \phasegate H^{\otimes 3}$. 
Here, $H$ denotes the single-qubit Hadamard gate. 
Successive applications of the Grover operator $\groverop = \reflectionop \oracleop$ onto $\ket{\psi}$ yield the final state $\groverop^r \ket{\psi}$ after $r$ repetitions. An optimal number $r$ exists for maximizing the probability of measuring one of the solution basis states~\cite{Boyer1998}. Representative quantum circuits can be found in the Supplemental Material (Figure~\ref{fig:circuit_grover_search} in Section~\ref{sec:quantum_circuits}).

A useful way to assess the performance of a Grover search is the algorithm success probability (ASP) of finding the oracle-tagged states and the squared statistical overlap (SSO) with the ideal probability distribution for the known target states~\cite{Figgatt2017}. 
For two probability vectors $\vec{p}$ and $\vec{q}$, we can determine their closeness through $\sso(\vec{p}, \vec{q}) = \left( \sum_k \sqrt{ p_k q_k} \right)^2$. 
As in Ref.~\cite{Figgatt2017}, the ASP is determined from the sum of all marked-state probabilities. 

\begin{figure*}
  \centering
  \includegraphics[width=\textwidth]{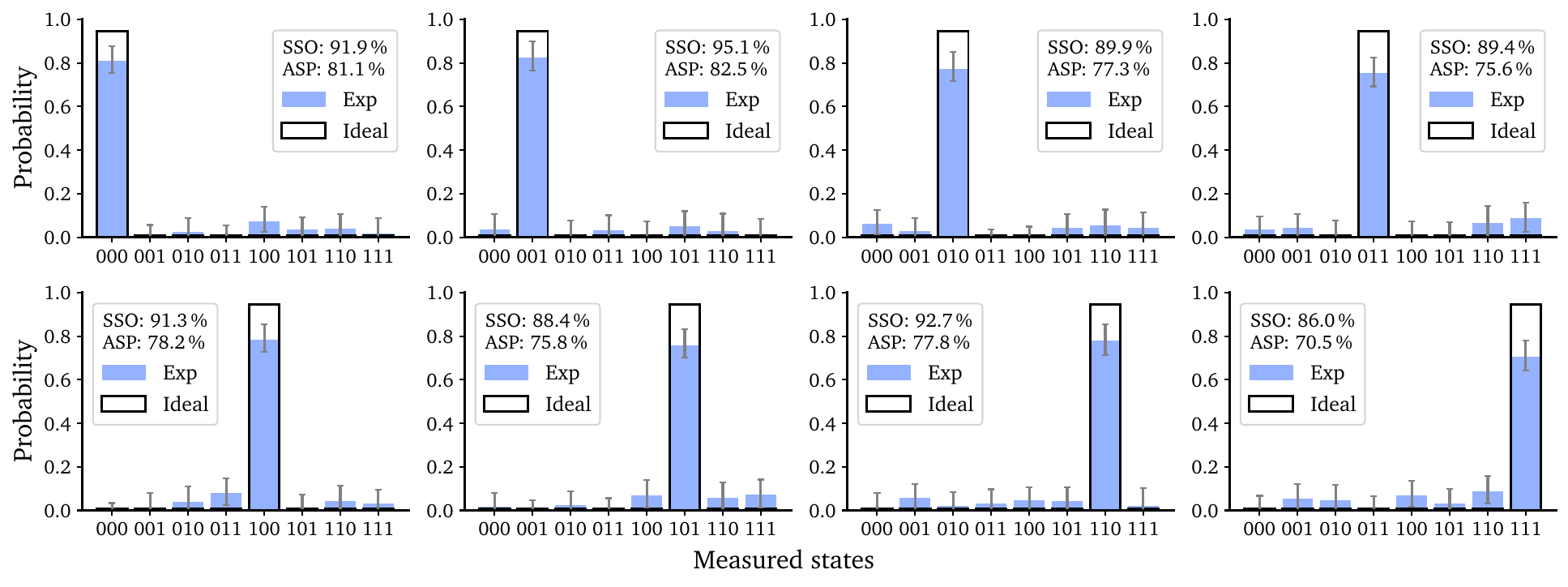}
  \caption{
    Grover search results for a single solution state and two Grover steps ($N=8$, $r=2$). The blue bars represent the measured experimental probabilities and the ideal distributions in black are shown for comparison. Each legend contains the individual squared statistical overlap and algorithm success probability.
  }
  \label{fig:result_grover_single_marked_states}
\end{figure*}

For a single target state, $\oracleop$ can be constructed from a single $\phasegate$ and local unitaries, e.g., $\oracleop = (X_0 X_2)\phasegate(X_0 X_2)$ for solution state $\ket{101}$ and Pauli operators $X_i$.
In addition, the search requires two applications of the Grover operator ($r=2$) for a maximum ideal ASP and, thus, needs four $\phasegate$ gates. 
Figure~\ref{fig:result_grover_single_marked_states} displays the measured Grover outcomes for all eight possible target states with an average ASP of $(77.3 \pm 3.4)\,\%$ alongside the ideal probability distributions with a maximum detection probability of $94.53\,\%$. 
The latter are used to determine an average SSO of $(90.6 \pm 2.6)\,\%$.
Although not entirely within striking distance to the best-performing average ASP of $89.4\,\%$ of this size at cryogenic temperatures ($T=15\,\text{mK}$) and with readout post-selection~\cite{Thorvaldson2025}, our results mark the highest achieved quantum search performance at ambient conditions for any quantum device to date. 
A detailed comparison to other modalities is presented in Table~\ref{tab:grover_other_modalities} and an extended version can be found in Table~\ref{tab:grover_other_modalities_extended} in Section~\ref{sec:literature_grover} of the Supplemental Material.
The maximum classical success probability for two oracle queries is 37.5\,\%~\cite{Rosmanis2024}, far below our reported quantum ASP.

\begin{figure}
  \centering
  \includegraphics[width=0.85\columnwidth]{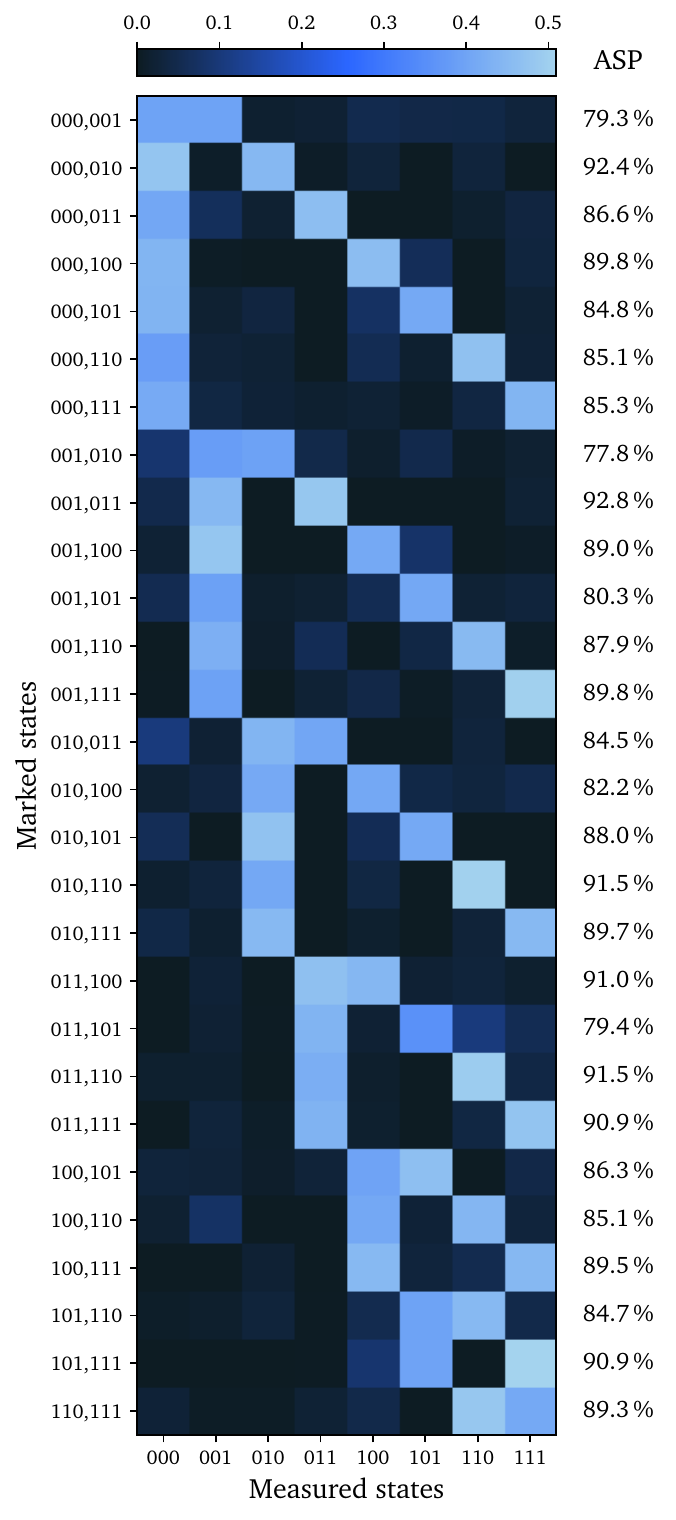}
  \caption{
      Measured probabilities for two oracle-tagged states after a single Grover step ($N=8$, $r=1$). All errors are of the same order as in Figure~\ref{fig:result_grover_single_marked_states} and, hence, omitted in this plot. Furthermore, all individual ASPs are located next to their corresponding row. 
      Since the ideal probabilities on the marked states are exactly 1/2, the SSO reads $(\sqrt{p_{1}} + \sqrt{p_{2}})^2/2$ and approaches $2p_1$, the ASP, for similar solution probabilities $p_2 \approx p_1$. 
      This situation matches the measured distributions and all SSOs are very close to the ASPs with differences less than 1\,\%. 
      For this reason, they are not shown here. 
  }
  \label{fig:result_grover_two_marked_states}
\end{figure}

If two target states are tagged by the oracle function, then one Grover step ($r=1$) suffices. The measured probabilities for all possible combinations are shown in Figure~\ref{fig:result_grover_two_marked_states} and, as before, demonstrate an excellent identification of the target states with an average ASP of $(87.0 \pm 4.2)\,\%$ and an almost identical mean SSO of $(86.9 \pm 4.2)\,\%$.
In comparison, the probability of classically finding one of the two marked states with a single oracle query is 46.4\,\%~\cite{Figgatt2017}.
\\

\begin{table*}[]
  \centering
  \begin{ruledtabular}
    \begin{tabular}{cccccccc}
      Modality & Repetitions $r$ & avg.~ASP & Sample temperature & Cryogenic & Vacuum & Remarks & Ref. \\
      \hline
      Neutral atoms & 1 & $\approx60\,\%$ & n.a. & $\times$ & \checkmark & \makecell{ensemble, only one target state,\\statistical post-processing} & \cite{Ahn2000} \\
      \hline
      \makecell{Liquid-state\\nuclear magnetic resonance} & 2 & $\approx65\,\%$ & $\sim297\,\text{K}$ & \checkmark & \checkmark & \makecell{ensemble, only one target state,\\very high magnetic field} & \cite{Vandersypen2000} \\
      \hline
      Superconducting & 1 & 49.2\,\% & 30\,\text{mK} & \checkmark & \checkmark & - & \cite{Roy2020} \\
       & 2 & 61.6\,\% & 15\,\text{mK} & \checkmark & \checkmark & - & \cite{Zhang2021} \\
       & 2 & 67.8\,\% & 15\,\text{mK} & \checkmark & \checkmark & - & \cite{Pokharel2024} \\
       & 2 & 78\,\% & 20\,\text{mK} & \checkmark & \checkmark & only one target state & \cite{Liu2025} \\
       & 2 & 68.1\,\% & n.a. & \checkmark & \checkmark & - & \cite{Leng2025} \\
      \hline
       Trapped ions & 1 & 43.7\,\% & n.a. & $\times$ & \checkmark & - & \cite{Figgatt2017} \\
       & 1 & 69\,\% & 6\,K & \checkmark & \checkmark & - & \cite{Shi2026} \\
      \hline
      Silicon spin qubit & 2 & 89.4\,\% & 15\,mK & \checkmark & \checkmark & Readout post-selection & \cite{Thorvaldson2025} \\
      \hline
      NV center & \textbf{2} & \textbf{77.3\,\%} & \textbf{296.3\,K} & $\times$ & $\times$ & - & \textbf{this work} \\
      \end{tabular}
  \end{ruledtabular}
  \caption{
    Published Grover search performances for search space $N=8$ and a single marked state on different quantum hardware. Results from our work are highlighted in bold.
  }
  \label{tab:grover_other_modalities}
\end{table*}

\emph{Device control.---}
Central to our quantum processor is a chip based on a type IIa, (100)-oriented, single crystal diamond substrate (ElementSix) with a natural 1.1\,\% abundance of \carbon{}, specified to less than 1 ppb of single nitrogen. 
Several surface preparation and annealing steps in conjunction with a sulphur implantation process facilitate the NV center creation and chip preparation.
The sulphur doping and activation was carried out according to Ref.~\cite{Luehmann2019} to increase the yield and stability of the NV centers.

The initialization of the electronic NV center spin is induced through incident light from a green laser (OBIS 514nm LX 30, Coherent) in a confocal microscope setup with $\text{NA}=0.95$ in air. This process cycles electronic population through a meta-stable state with non-spin-conserving transitions and effectively polarizes the electronic spin in its $\kete{m_S=0}$ state~\cite{Doherty2013}.

An inherent zero-field splitting of the NV center spin separates the $\kete{m_S=0}$ from $\kete{m_S=\pm 1}$ and an additional constant magnetic field aligned along the NV symmetry axis further separates all three states energetically. Tuning this magnetic field to approximately $\sim$51.6\,mT allows us to address all allowed spin transitions selectively and polarizes the nitrogen and carbon spins under long laser illumination~\cite{Jacques2009, Smeltzer2011, Dreau2012}. 
Control over the electronic and nuclear spin transitions is exerted through microwave and radiofrequency pulses from an arbitrary waveform generator, which are subsequently amplified.
An avalanche photodiode and single photon counting are used for the detection of fluorescent light emitted from the NV center after laser shots.
Depending on the electronic spin states, the transient intensity varies and is used to infer the state population~\cite{Steiner2010}.
As our system works at ambient conditions, 
we compensate externally induced system drifts by means of automatic correction procedures.
Typical external thermal variations during the measurements for this work were in a temperature interval of $\Delta T \approx 0.8\,\text{K}$ at average temperature $T \approx 296.3\,\text{K}$.
\\

After initialization of the electronic spin to $\kete{m_S=0}$, the transition frequency for $\kete{m_S=0} \leftrightarrow \kete{m_S=-1}$ conditional on the nuclear state $\ket{000}$ was detected through optically detected magnetic resonance. 
A robust electronic spin population inversion is achieved through adiabatic pulses~\cite{Spindler2016}, for which we subsequently calibrate an optimal amplitude and pulse length, achieving a pseudo-fidelity greater than 99\,\% (Figure~\ref{fig:1q_rb}). 
The inversion to $\kete{m_S=-1}$ causes an additional energy splitting for the nuclear spins through hyperfine couplings and spectrally separates the \carbon{} spins.
These nuclear transition frequencies can be detected via electron–nuclear double resonance (ENDOR) spectroscopy~\cite{Yavkin2016}, which probes the system with varying radiofrequency pulses.
In order to ensure an adiabatic evolution of the electronic state, the flat-top pulses are appropriately tapered at the leading and trailing edges.
\begin{table}[b]
  \centering
  \begin{ruledtabular}
    \begin{tabular}{cccc}
      Nuclear spin & Frequency [MHz] & Rabi frequency [kHz] & $T_2^*\,[\text{ms}]$ \\
      \hline
      \nitrogen & 2.940059(30) & 30.96(2) & $\approx 4.3$ \\
      $\carbonmath_{1}$ & 5.990373(69) & 52.33(8) & $\approx 1.5$ \\
      $\carbonmath_{2}$ & 13.314576(63) & 108.61(6) & $\approx 1.4$ \\
      \end{tabular}
  \end{ruledtabular}
  \caption{
    Nuclear spin gate parameters and Ramsey coherence times. Numerical values in parentheses denote the error in the last digits.
  }
  \label{tab:nuclear_parameters}
\end{table}
Measured ENDOR data and fitting results are provided in Figure~\ref{fig:endor} in Section~\ref{sec:endor} of the Supplemental Material.
Once the nuclear frequencies are approximately known, estimates for the correct pulse lengths are obtained from nuclear Rabi experiments.

Thereafter, the radiofrequency pulse parameters are further fine-tuned by utilizing two experimental sequences from deterministic benchmarking~\cite{Tripathi2024, Tripathi2025}. 
In particular, we iteratively measure the probabilities $P_{YY} = \abs*{ \mel{+}{\left[ R_y(\pi) R_y(\pi) \right]^n }{+} }^2$ and $P_{X\widebar{X}} = \abs*{ \mel{+}{\left[ R_x(\pi) R_x(-\pi) \right]^n }{+} }^2$ for varying pulse length and frequency, respectively. 
Gate times for $\pi$-rotations are denoted by $t_g$. 
For each increase in $n$, the pulse parameters from the previous calibration are used as guiding mean value and new parameters are determined from the maximum probability position within the time window $\delta t = t_g /(2n)$ and frequency window $\delta\omega = \pi / (8t_g n)$, respectively. 
This procedure allows us to calibrate the pulse frequency up to an uncertainty of $\sim$60\,Hz, which is on the order of the thermal fluctuations of the resonance frequencies of strongly coupled nuclear spins~\cite{Xu2023}. Table~\ref{tab:nuclear_parameters} provides an overview for nuclear gate parameters and coherence times.
With the help of these optimized pulses, a spin polarization transfer protocol enhances the initial nuclear spin polarization as shown in Figure~\ref{fig:odmr_comparison_init_sequence} in Section~\ref{sec:nuclear_polarization} of the Supplemental Material.

The final calibration concerns the multi-qubit $\phasegate$ gate implementation. 
We optimize the corresponding microwave pulse shape through numerical simulations including hyperfine tensors from a public dataset~\cite{Takacs2024}, which are adjusted to match experimentally determined values. 
As a result, we achieve a gate time of just 810\,ns. Any remaining unwanted nuclear phases are corrected through suitable virtual $z$-rotations, which are determined from SPAM-robust phase measurements through repetitive application of the uncorrected $\phasegate$.

At last, the state readout proceeds through all $2^3$ corresponding electronic transitions. 
A non-negative least square procedure and subsequent normalization yield the measurement probabilities $\propabilityvector \geq 0$ from the linear relationship $\intensityvector = M \propabilityvector$, where $\intensityvector$ denotes the measured fluorescence. 
The matrix $M$ is measured in advance for all eight basis states. 
\\

\emph{Conclusion and outlook.---}
Running quantum computations outside of a precisely controlled lab environment poses a major challenge. 
Here, we report the highest Grover search success rate with three qubits among any quantum processors at ambient conditions to date~(Table~\ref{tab:grover_other_modalities}).
Based on an NV center in diamond, we achieve an average success probability of 77.3\,\% and 87.0\,\% for single and two marked states, respectively. Both values lie significantly higher than their highest achievable classical counterparts with 37.5\,\% and 46.4\,\%.
Noteworthy and testament to the absence of cryogenics and vacuum equipment is also the low operational power consumption of roughly 600\,W for our quantum computer.

In future experiments, a better fidelity estimation of $\phasegate$ can be achieved through newer benchmarking approaches~\cite{Korliakov2025, Ye2026}, addressing the full dimensionality. 
Moreover, the pulse quality is currently limited by a finite amplitude resolution of the waveform generator and, based on simulations, we expect higher fidelities of at least 99\,\% with a more precise pulse generation.
\\

\emph{Acknowledgements---}
This research is co-financed with tax revenues on the basis of the budget approved by the Saxon State Parliament as part of the SAX-QT project.


\bibliography{references}


\clearpage

\onecolumngrid

\begin{center}
{\large\bfseries Supplemental Material --- Grover Search with Semiconductor Spin Qubits at Ambient Conditions}\\[1ex]

{\normalsize Sebastian~Gemsheim\textsuperscript{1}, Fabian~Kl\"upfel\textsuperscript{1}, Max~Knei\ss\textsuperscript{1}, Nicole~Raatz\textsuperscript{1}, Tobias~Herzig\textsuperscript{1}, Matthias~Mendt\textsuperscript{1}, Evgeny~Kreissig\textsuperscript{1}, Ulrike~R\"uckert\textsuperscript{2}, Albrecht~H\"anel\textsuperscript{2}, Jan~Meijer\textsuperscript{1,3}, and Marius~Grundmann\textsuperscript{1,3}\\
\vspace{2mm}
\small
\textit{\textsuperscript{1}SAXON Q GmbH, Emilienstraße 15, 04107 Leipzig, Germany}\\
\textit{\textsuperscript{2}Fraunhofer IWU, N\"othnitzer Straße 44, 01187 Dresden, Germany}\\
\textit{\textsuperscript{3}Felix-Bloch-Institute for Solid State Physics, University of Leipzig, Linn\'estraße 5, 04103 Leipzig, Germany}
}
\end{center}


\setcounter{secnumdepth}{2}

\setcounter{section}{0}
\setcounter{equation}{0}
\setcounter{figure}{0}
\setcounter{table}{0}

\renewcommand{\thesection}{S\arabic{section}}
\renewcommand{\theequation}{S\arabic{equation}}
\renewcommand{\thefigure}{S\arabic{figure}}
\renewcommand{\thetable}{S\arabic{table}}

\renewcommand{\theHsection}{supp.\arabic{section}}
\renewcommand{\theHequation}{supp.\arabic{equation}}
\renewcommand{\theHfigure}{supp.\arabic{figure}}
\renewcommand{\theHtable}{supp.\arabic{table}}


\section{Spectral signature of nuclear spins}
\label{sec:endor}
\begin{figure}[h]
    \centering
    \includegraphics[width=0.45\columnwidth]{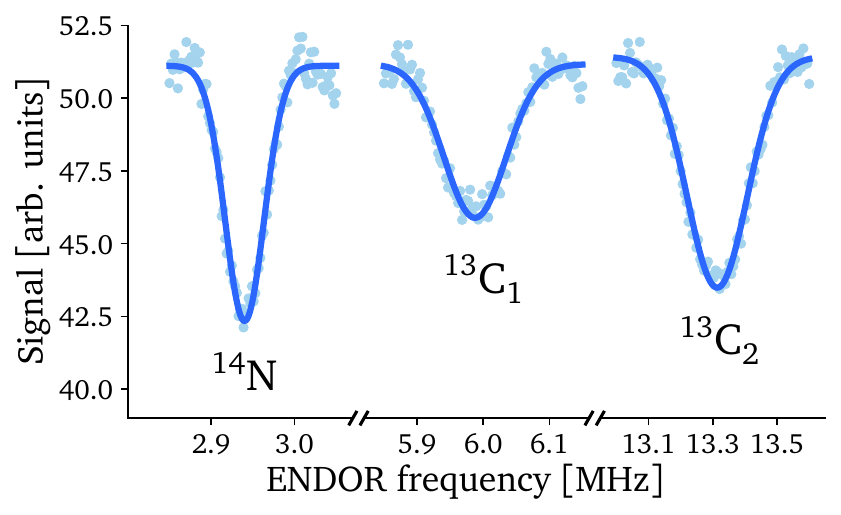}
    \caption{
        ENDOR data for electronic state manifold $m_S=-1$. Experimental data points are fit by a Gaussian curve $A - B e^{-(f-f_0)^2/\sigma^2}$ (blue) and provide the center frequencies estimates $f_0[\text{\nitrogen{}}]=2.940\,\text{MHz}$, $f_0[{}^{13}\text{C}_{1}]=5.988\,\text{MHz}$ and $f_0[{}^{13}\text{C}_{2}]=13.314\,\text{MHz}$. Moreover, the widths for each dip read $\sigma[\text{\nitrogen{}}]\approx 33\,\text{kHz}$, $\sigma[{}^{13}\text{C}_{1}]\approx 66\,\text{kHz}$ and $\sigma[{}^{13}\text{C}_{2}]\approx 132\,\text{kHz}$.
    }
    \label{fig:endor}
\end{figure}

Electron–nuclear double resonance spectroscopy (ENDOR) allows us to detect the nuclear spin qubits surrounding the NV center. The measured responses for spectrally varying radio-frequency pulses is shown in Figure~\ref{fig:endor}. The spectral signature of the measured carbon spins are known~\cite{Dreau2012} and can be used to estimate their distances to the NV center, which are between $2.5$ and $4\,\mathring{A}$.


\section{Efficiency of nuclear spin polarization sequences}
\label{sec:nuclear_polarization}
\begin{figure}[h]
    \centering
    \includegraphics[width=0.5\columnwidth]{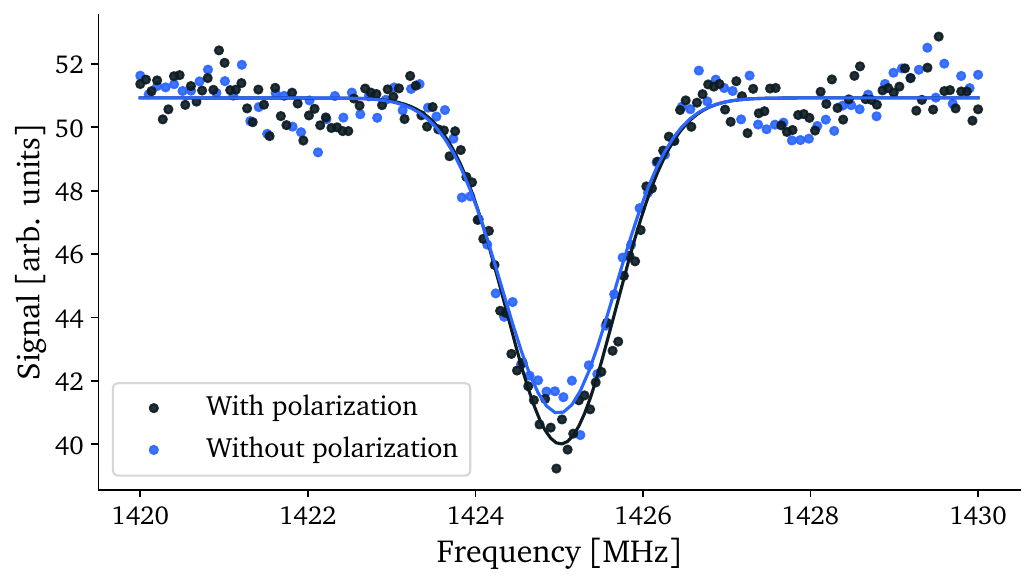}
    \caption{
        Optically detected magnetic resonance spectra with and without additional polarization transfer from the elctron to the nuclear spins. Gaussian fits are used for a better visual distinction.
    }
    \label{fig:odmr_comparison_init_sequence}
\end{figure}

At the excited-state anti-crossing, at around 50\,mT, not only the nitrogen spin of the NV center polarizes under laser illumination, but also some strongly coupled \carbon{} spins~\cite{Smeltzer2011, Dreau2012}. Their degree of polarization can be further increased by additionally utilizing a polarization transfer protocol~\cite{Waldherr2014}.
In order to gauge the effectiveness of this procedure, we compare optically detected magnetic resonance spectra with and without the transfer routines in Figure~\ref{fig:odmr_comparison_init_sequence}. Clearly, the achievable contrast increases by about 10\,\% to a total of 21.4\,\% and indicates an improved nuclear spin polarization.

\section{Quantum circuits for Grover search}
\label{sec:quantum_circuits}
\begin{figure}[h]
    \includegraphics[height=32mm]{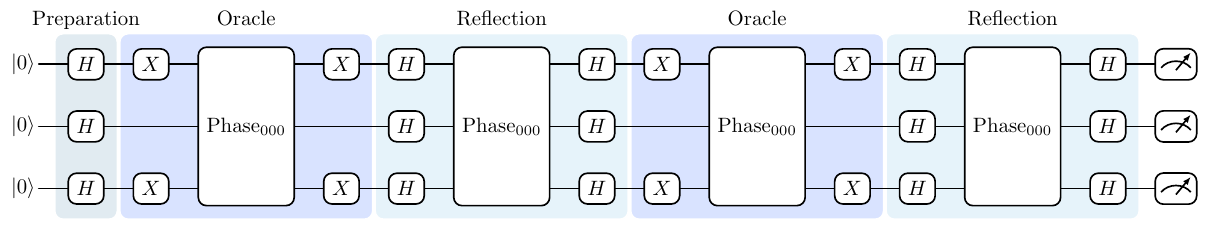}
    \includegraphics[height=32mm]{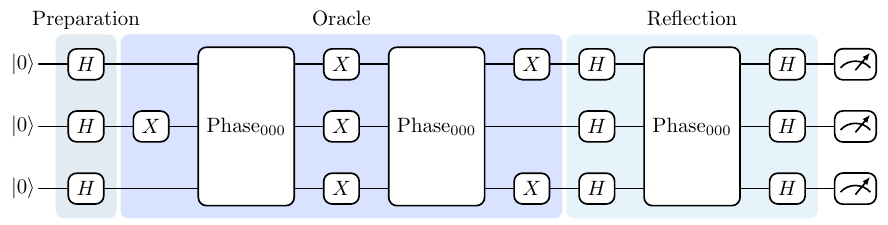}
    \caption{
        Circuit representations of the Grover seach algorithm for three qubits ($N=8$). The upper plot shows two consecutive applications ($r=2$) of the Grover operator $G$ for the marked state $\ket{101}$. For two marked state (e.g., $\ket{010}$ and $\ket{101}$), the oracle requires two $\phasegate$ gates, but one Grover step ($r=1$) suffices for a maximum success probability (lower plot). 
    }
    \label{fig:circuit_grover_search}
\end{figure}

The main text only features the the unitary operator for the corresponding Grover search. Here, we additionally show the associated quantum circuits representing the decomposition in our native gate set in Figure~\ref{fig:circuit_grover_search} for an exemplary oracle setting.

\section{Grover results in literature}
\label{sec:literature_grover}
Grover searches have been experimentally implemented on several quantum devices. While some are given in the main text, we extend the comparison of our work to available literature for arbitrary search space dimensions. Without claiming completeness, Table~\ref{tab:grover_other_modalities} shows an extended list of state-of-the-art Grover results for different modalities.
\begin{sidewaystable}[h]
  \centering
  \begin{ruledtabular}
    \begin{tabular}{ccccccccc}
      Modality & Search space $N$ & Repetitions $r$ & avg.~ASP & Sample Temperature & Cryogenic & Vacuum & Remarks & Ref. \\
      \hline
      Neutral atoms & 8 & 1 & $\approx60\,\%$ & n.a. & $\times$ & \checkmark & \makecell{ensemble,\\only one target state,\\statistical post-processing} & \cite{Ahn2000} \\
      \hline
      Liquid-state nuclear magnetic resonance & 8 & 2 & $\approx65\,\%$ & $\sim297\,\text{K}$ & \checkmark & \checkmark & \makecell{ensemble,\\only one target state,\\very high magnetic field} & \cite{Vandersypen2000} \\
      \hline
      Superconducting & 4 & 1 & 81.1\,\% & $13\,\text{mK}$ & \checkmark & \checkmark & - & \cite{DiCarlo2009} \\
       & 8 & 1 & 49.2\,\% & 30\,\text{mK} & \checkmark & \checkmark & - & \cite{Roy2020} \\
       & 8 & 2 & 61.6\,\% & 15\,\text{mK} & \checkmark & \checkmark & - & \cite{Zhang2021} \\
       & 16 & 2 & 16.3\,\% & 15\,\text{mK} & \checkmark & \checkmark & - & \cite{Zhang2021} \\
       & 32 & 1 & 3.9\,\% & 15\,\text{mK} & \checkmark & \checkmark & - & \cite{Zhang2021} \\
       & 32 & 1 & 4.4\,\% & 15\,\text{mK} & \checkmark & \checkmark & only one target state & \cite{Zhang2022} \\

       & 32 & 2 & 3.3\,\% & 15\,\text{mK} & \checkmark & \checkmark & only one target state & \cite{Zhang2022} \\
       & 16 & 1 & 34.2\,\% & 10\,\text{mK} & \checkmark & \checkmark & - & \cite{Chu2022} \\
       & 64 & 1 & 3.9\,\% & 10\,\text{mK} & \checkmark & \checkmark & - & \cite{Chu2022} \\
       & 9 & 1 & 44.4\,\% & 15\,\text{mK} & \checkmark & \checkmark & - & \cite{Roy2023} \\
       & 9 & 2 & 49.6\,\% & 15\,\text{mK} & \checkmark & \checkmark & - & \cite{Roy2023} \\
       & 8 & 2 & 67.8\,\% & 15\,\text{mK} & \checkmark & \checkmark & - & \cite{Pokharel2024} \\
       & 16 & 2 & 38.1\,\% & 15\,\text{mK} & \checkmark & \checkmark & - & \cite{Pokharel2024} \\
       & 32 & 2 & 15.3\,\% & 15\,\text{mK} & \checkmark & \checkmark & - & \cite{Pokharel2024} \\
       & 8 & 2 & 78\,\% & 20\,\text{mK} & \checkmark & \checkmark & only one target state & \cite{Liu2025} \\
       & 8 & 1 & 62.0\,\% & n.a. & \checkmark & \checkmark & - & \cite{Leng2025} \\
       & 8 & 2 & 68.1\,\% & n.a. & \checkmark & \checkmark & - & \cite{Leng2025} \\
      \hline
      Trapped ions & 32 & 1 & 22.3\,\% & 12.6\,K & \checkmark & \checkmark & \makecell{Quantinuum,\\only one target state} & \cite{Zhang2022} \\
       & 32 & 2 & 49.0\,\% & 12.6\,K & \checkmark & \checkmark & \makecell{Quantinuum,\\only one target state} & \cite{Zhang2022} \\
       & 32 & 1 & 7.4\,\% & 288\,K & $\times$ & \checkmark & \makecell{IonQ,\\only one target state} & \cite{Zhang2022} \\
       & 32 & 2 & 4.0\,\% & 288\,K & $\times$ & \checkmark & \makecell{IonQ,\\only one target state} & \cite{Zhang2022} \\
       & 8 & 1 & 43.7\,\% & n.a. & $\times$ & \checkmark & - & \cite{Figgatt2017} \\
       & 4 & 1 & 71\,\% & $\sim$297\,K & $\times$ & \checkmark & distributed & \cite{Main2025} \\
       & 8 & 1 & 40\,\% & n.a. & $\times$ & \checkmark & \makecell{error corrected,\\ one instance of two target states} & \cite{Butt2026} \\
       & 5 & 1 & 96.8\,\% & 6\,K & \checkmark & \checkmark & - & \cite{Shi2026} \\
       & 8 & 1 & 69\,\% & 6\,K & \checkmark & \checkmark & - & \cite{Shi2026} \\
      \hline
      Photonic & 4 & 1 & 96.4\,\% & n.a. & $\times$ & $\times$ & MBQC & \cite{Ciampini2016} \\
       & 4 & 1 & 98.7\,\% & n.a. & n.a. & n.a. & MBQC & \cite{Feng2026} \\
       & 4 & 1 & 80.8\,\% & n.a. & \checkmark & \checkmark & MBQC & \cite{Huster2026} \\
      \hline
      Silicon spin qubit & 4 & 1 & 75.3\,\% & 20\,mK & \checkmark & \checkmark & - & \cite{Watson2018} \\
       & 4 & 1 & 96.8\,\% & 20\,mK & \checkmark & \checkmark & - & \cite{Noiri2022} \\
       & 8 & 2 & 89.4\,\% & 15\,mK & \checkmark & \checkmark & Readout post-selection & \cite{Thorvaldson2025} \\
      \hline
      NV center & 4 & 1 & 85.5\,\% & $\sim$297\,K & $\times$ & $\times$ & - & \cite{vanderSar2012} \\
       & 4 & 1 & 83\,\% & $\sim$297\,K & $\times$ & $\times$ & - & \cite{Wu2019} \\
       & 4 & 1 & 84.6\,\% & $\sim$297\,K & $\times$ & $\times$ & - & \cite{Zhang2020} \\
       & \textbf{8} & \textbf{2} & \textbf{77.3\,\%} & \textbf{296.3\,K} & $\times$ & $\times$ & - & \textbf{this work} \\
      \end{tabular}
  \end{ruledtabular}
  \caption{
    Literature Grover results for different modalities and their experimental requirements. Here, measurement-based quantum computation~\cite{Briegel2009} is abbreviated by MBQC. Results from our work are highlighted in bold.
  }
  \label{tab:grover_other_modalities_extended}
\end{sidewaystable}

\end{document}